\documentstyle{article}
  \begin{document}
  \title{Possible mechanism of electrical field origin around
  celestial bodies }
  \author{G.S. Bisnovatyi-Kogan
  \thanks{Institute of Space Research (IKI), Moscow, Russia, Profsoyuznaya
  84/32, Moscow 117810, Russia,  gkogan@mx.iki.rssi.ru; and Theoretical Astrophysics Center (TAC), Copenhagen}}
  \date{}
  \maketitle
  \begin{abstract}
  Slow magnetic field variations in stars and planets create a
quasistationary electrical field which may be observed. It is supposed that the electrical field near the Earth surface may be partially connected with variation of the  Earth magnetic field. Two examples of the electrical field distribution around the infinite cylinder, and the circular loop with a lineary growing with time electrical currents are given.
\end{abstract}

\smallskip


\smallskip

There is a vertical negative electrical field near the Earth surface of the order of
100 V/m $\sim 3\cdot 10^{-3}$ CGSE which origin is not quite clear, but which usually is associated with the action of thunderstorms. The Earth is negatively charged, and the electron flux into the Earth is balanced by transport of the positive charge from the Earth into the atmosphere clouds. Not only the causes, but also the effects of electrification of clouds are still far from being understuud  (Vonnegut, 1973). The electrical field of the Eath is fluctuating, and horizontal component may also appear.

Electrical field induced by rotating Earth dipole $E_d$ is very small

\begin{equation}
\label{edip}
E_d\sim \frac{v_r}{c} B \sim 2 \cdot 10^{-7} CGSE
\end{equation}
for the surface rotational velocity $v_r\approx 6\cdot 10^3$cm/s, and $B \sim 1$ Gauss. In this note we show that the induced electrical field may be much larger than the rotationally induced one, if it is connected with nonstationary currents in the Earth conducting cores, which produce much larger magnetic fields inside the Earth.

If the current loop is surrounded by matter with negligible conductivity, the magnetic field produced outside of such loop may be calculated like it is surrounded by vacuum. The electrical field in the highly conducting media is small, so we may use the boundary condition of zero electrical field at the surface of the current loop. The problem is solved in a very simple way for a lineary growing curent. If the electrical current is concentrated in the torus with the large radius $a$, small radius $b\ll a$, than near the Earth surface the ration of the electrical and magnetic fields induced by the same current loop is of the order (see (\ref{ac7}),(\ref{comp}))
$\frac{E_{\phi}}{|B|} \sim \ln\left(\frac{8a}{b}\right) \frac{r^3}{a^2 ct}$.
Taking the value 10 for the logarithm, $r/a \sim 10$, and 10 days for the characteristic time, we obtain for the Earth radius $r=6\cdot 10^8$ cm and Earth poloidal magnetic field $\sim 1$ Gauss the induced horizontal electrical field of the order of $2\cdot 10^{-5}$ Gauss. It is far from the values of the vertical electrical field, but much exceeds the value of the electrical field induced by the rotating dipole. On the planets where the thunderstorm activity is surpressed the electrical field component produced by the variable electrical current inside the planet may become prevailable.

       Let us consider configuration of electrical fields around infinite cylinder with an electrical current along its axis, and around a current loop. In both cases the electrical current is lineary growing with time.
Start from Maxwell equations 

\begin{equation}
\label{maxh}
{\rm div}\,{\bf H}=0,
\quad
{\rm rot}\,{\bf H}=\frac{1}{c} \frac{\partial {\bf E}}{\partial t}
 + \frac{4\pi}{c} {\bf j},
\end{equation}

\begin{equation}
\label{maxe}
{\rm rot}\,{\bf E}=-\frac{1}{c} \frac{\partial {\bf H}}{\partial t},
\quad
{\rm div}\,{\bf E}=4\pi \rho_e.
\end{equation}
For linear dependence of the current density ${\bf j}$ and magnetic field strength ${\bf H}$ on time 

\begin{equation}
\label{time}
{\bf j}={\bf i}\, t \qquad
{\bf H}={\bf h}\, t
\end{equation}
the electrical field does not depend on time with zero charge density $\rho_e$. 
Inside the matter it corresponds 
to constant
divergentless external electromoving force $E_{ext}$, which is balanced by the electrical field in the perfect plasma with infinite conductivity, considered here. With account of (\ref{time}) Maxwell equations  (\ref{maxh}),(\ref{maxe}) 
reduce to

\begin{equation}
\label{maxth}
{\rm div}\,{\bf h}=0,
\quad
{\rm rot}\,{\bf h}= \frac{4\pi}{c} {\bf i},
\end{equation}

\begin{equation}
\label{maxte}
{\rm rot}\,{\bf E}=-\frac{{\bf h}}{c},
\quad
{\rm div}\,{\bf E}=0.
\end{equation}
Introducing vector-potential ${\bf A}={\bf a}t$, such as

\begin{equation}
\label{vec}
{\rm rot}\,{\bf a}={\bf h},
\quad
\Delta {\bf a}= -\frac{4\pi}{c} {\bf i},
\end{equation}
we have from comparison of (\ref{vec}) with (\ref{maxth}) the relation

\begin{equation}
\label{ae}
{\bf E}=-\frac{\bf a}{c}.
\end{equation}
The vector-potential is not defined uniquely, but the electrical field is 
defined uniquely with account of boundary conditions. Cosider two particular cases.
 
{\bf 1. Infinite curent along $z$-axis}. In cylindrical coordinate system
$(r, \phi, z)$ $\partial/\partial\phi=\partial/\partial z=0$, and
there is only one non-zero component $a_z$, satisfying equation

\begin{equation}
\label{az}
\frac{1}{r}\frac{\partial}{\partial r}
\left(r\frac{\partial a_z}{\partial r}\right) = -\frac{4\pi}{c} i_z.
\end{equation}
 The solution of this equation is

\begin{equation}
\label{az1}
h_{\phi}=-\frac{da_z}{dr}=\frac{2I_z(r)}{cr}, \quad {\rm at} \,\,\, r<b,
\quad h_{\phi}=\frac{2I_z(b)}{cr}, \quad {\rm at} \,\,\, r>b,
\end{equation}
$$\,\,\,I(r)=2\pi\int_0^r{i(x) dx},\,\,\, J=It ; $$
$$a_z=-\frac{2}{c}\int_0^r{\frac{I_z(x)dx}{x}}
+C_1,  \quad {\rm at} \,\,\, r<b,
\quad  a_z=-\frac{2}{c}I(b)\ln r +C_2,  \quad {\rm at} \,\,\, r>b.$$
For continous $a_z$ there is only one arbitrary constant. The choice of this constant in the expression for $E$ depends on the boundary condition on the boundary $r=b$, which, in turn, is defined by a distribution of $E_{ext}$. Taking $E(b)=-E_{ext}(b)=0$ we obtain for the distribution of $E$ outside the conducting cylinder in the form

\begin{equation}
\label{az2}
 E_z=\frac{2}{c^2}I(b)\ln\frac{r}{b}.
\end{equation} 
So, the linear infinitely long current, lineary growing with time, produces around a stationary electrical field parallel to itself, growing logarithmically with a distance from the current.

{\bf 2. Circular current.} Consider a circular current of the radius $a$ around the coordinate center in the plane $z=0$. We are interested in the solution in vacuum around this circuit, which may be obtained using the solution for a constant current in Landau and Lifshits (1993). For a lineary growing current the vector potential with the only non-zero component $A_{\phi}=a_{\phi}t$ has a form in cylindrical coordinates $(r,\,\phi,\, z)$

\begin{equation}
\label{ac1}
a_{\phi}=\frac{4I}{ck}\sqrt{\frac{a}{r}}[(1-\frac{1}{2}k^2)K-E],
\end{equation}
where $k$ and complete elliptic integrals of the first and second kinds $K$ and $E$ are expressed as

\begin{equation}
\label{ac2}
k^2=\frac{4a r}{(a+r)^2+z^2}, 
\end{equation}
$$K=\int_0^{\pi/2}{\frac{d\theta}{\sqrt(1-k^2\sin^2\theta)}}, \,\,\,
  E=\int_0^{\pi/2}{\sqrt(1-k^2\sin^2\theta)}d\theta.
$$
The electrical field is determined from (\ref{ac1}) with an arbitrary function, obtained from boundary conditions, which reduses to a constant in this case:

\begin{equation}
\label{ac3}
E_{\phi}=-\frac{4I}{c^2k}\sqrt{\frac{a}{r}}[(1-\frac{1}{2}k^2)K-E]+D.
\end{equation}
The constant $D$ is obtained from the boundary conditions on the surface of the ring current, which has a form of torus with a very small cross-section radius $b\ll a$. The equation of the torus surface and the value of elliptic integrals on it with $b \ll a$ are (Gradshtein, Ryzhik, 1996)

$$
(r-a)^2+z^2=b^2, \quad k^2\approx 1-\frac{(r-a)^2+z^2}{4a^2}=1-\frac{b^2}{4a^2},
$$
\begin{equation}
\label{ac4}
 K\approx \ln\frac{4}{\sqrt{1-k^2}}=\ln\frac{8a}{b}, \quad E\approx 1,
\end{equation}
$$E_{\phi s}=-\frac{2I}{c^2}\left(\ln\frac{8a}{b}-2\right)+D.$$
Taking $E_{\phi s}=0$ we finally obtain the electrical field distribution around the torus-like circuit current as

\begin{equation}
\label{ac5}
E_{\phi}=\frac{2I}{c^2}\left(\ln\frac{8a}{b}-2-
\frac{2}{k}\sqrt{\frac{a}{r}}[(1-\frac{1}{2}k^2)K-E]\right).
\end{equation}
At large distance from the torus at $r^2+z^2 \gg a^2,\,\, k\ll 1$ we find the electrical field, using $E$ and $K$ expansions, in the form
\begin{equation}
\label{ac6}
E_{\phi}=\frac{2I}{c^2}\left(\ln\frac{8a}{b}-2-
\frac{\pi a^2r}{2(r^2+z^2)^{3/2}}\right), \quad k^2\approx
\frac{4ar}{r^2+z^2}\left(1-\frac{2ar}{r^2+z^2}\right),
\end{equation}
$$E=\frac{\pi}{2}\left(1-\frac{k^2}{4}-\frac{3k^4}{64}\right), \quad
K=\frac{\pi}{2}\left(1+\frac{k^2}{4}+\frac{9k^4}{64}\right).
$$
Write also the magnetic field components at large distances

\begin{equation}
\label{ac7}
B_r=-\frac{\partial A_{\phi}}{\partial z}=-\frac{3\pi a^2 It}{c}
\frac{zr}{(r^2+z^2)^{5/2}},
\end{equation}
$$
B_z=\frac{1}{r}\frac{\partial(r A_{\phi})}{\partial r}=\frac{\pi a^2 It}{c}
\frac{2z^2 -r^2}{(r^2+z^2)^{5/2}},
$$
and on $z$-axis at $r=0$
\begin{equation}
\label{ac8}
B_r=0,\quad B_z=\frac{\pi a^2 It}{c} \frac{2}{z^3}, \quad E_{\phi}=0.
\end{equation}
The solution (\ref{ac5}), as well as (\ref{az2}), are obtained for a linear time dependence of the electrical current at all $t>-\infty$, and corresponds to infinite electrostatic energy $\int(E^2/8\pi)dV$. In a realistic case when the electrical current is started at $t=0$ the solution (\ref{ac5}) is valid only in the region $r<ct$, while outside we still have $E_{\phi}=0$.

Note that at $a \rightarrow \infty$ the solution (\ref{ac5}) tends to (\ref{az2}). This limit should be done in the coordinate system with the origin in the point $(a,0,0)$, where $x$-axis intersects with the circle.
In this system

\begin{equation}
\label{ac9}
r=\sqrt{(x-a)^2+y^2}, \quad 
k^2=\frac{4a\sqrt{(x-a)^2+y^2}}{(a+\sqrt{(x-a)^2+y^2})^2+z^2}.
\end{equation}
In the limit $a \rightarrow \infty$ we obtain from (\ref{ac9}),(\ref{ac4}), (\ref{ac5})

\begin{equation}
\label{ac10}
 k^2\approx 1-\frac{x^2+z^2}{4a^2}, \quad r^2=x^2+z^2,\quad
K(k)\approx \ln{\frac{8a}{r}}, \quad E_{\phi} \approx \frac{2I}{c^2}\ln\frac{r}{b},
\end{equation}
what coincides exactly with (\ref{az2}).

The ratio $E_{\phi}/|B|$ at large distances $z\sim r \gg a$ is of the order of

\begin{equation}
\label{comp}
\frac{E_{\phi}}{|B|} \sim \left(\ln\frac{8a}{b}-2\right) \frac{r^3}{a^2 ct}.
\end{equation}

\medskip

{\Large{\bf Acknowledgement}}

\medskip

Author is grateful to Theoretical Astrophysics Center (TAC), Copenhagen, for
hospitality during the work on this paper.


\begin{thebibliography}{}

\bibitem{gr} Gradshtein I.S, Ryzhik I.M. 1996, Tables of integrals, series and products. Academic Press, INC.
\bibitem{ll} Landau L.D., Lifshitz E.M. 1993,  Electrodynamics of continous     media. Pergamon Press.
\bibitem{von} Vonnegut B. 1973, Ann. Rev. Earth and Planet. Sci., {\bf 1}, 297.

\end{thebibliography}
\end{document}